# About oscillation method to study liquid viscosity

**I.H. Umirzakov**
Novosibirsk, Russia
e-mail: cluster125@gmail.com

## Abstract

It is shown that the oscillation method to study liquid viscosity of [1-3,7-34] is based on incorrect consideration of one dimensional forced (constrained) )vibrations of plate in viscous liquid, because additional (apparent) mass of liquid and surface forces are incorrectly taken into account, and Archimedes force did not taken into account. It is shown the vibrations cannot be one dimensional, and torsional vibrations need to be take into account, the results obtained with the use of the method are not reliable and incorrect.

УДК 532.1

# Вибрационный метод измерения вязкости жидкости

И.Х. Умирзаков
г. Новосибирск, Россия
e-mail: cluster125@gmail.com

## Аннотация

Показано, что вибрационный метод измерения вязкости жидкости, предложенный и использованный в [1-3,7-34], являющийся основой вибрационного метода фазового анализа [7-34], основан на неправильном решении одномерных колебаний тонкой пластинки в вязкой жидкости, неправильно учтены присоединенная масса и поверхностные силы, не учтена сила Архимеда. Показано, что колебания не могут быть одномерными, отвечающими колебаниям вдоль оси трубки, нельзя пренебречь крутильными колебаниями, колебания не могут быть гармоническими, характеристик колебательной системы не являются постоянными во времени, в системе не может существовать метастабильная жидкость и поэтому с помощью колебательной системы невозможно изучение процесса затвердевания и плавления одно- и многокомпонентных веществ, в системе не достигаются изотермические условия. Приведены также другие доказательства неприменимости этого вибрационного метода измерения вязкости жидкости.

# Введение

Вибрационный метод измерения вязкости жидкости основан на задаче о малых вынужденных колебаний тонкой пластинки толщины $d$, жестко связанной с тонкой трубкой малого диаметра, подвешенной на пружине, верхний конец которой закреплен, и совершающей гармонические колебания под действием внешней гармонической силы $f \cdot \cos(\omega \cdot t)$ с амплитудой $f = const$ с частотой $\omega = const$, и в определении произведения $\eta \cdot \rho$ вязкости жидкости $\eta$ на ее плотность $\rho$ через фазу $\varphi$ и амплитуду колебаний пластинки $A$ [1-3]. Пластинка и нижняя часть трубки погружены в жидкость. Колебания происходят в вертикальной плоскости, проходящей через всю поверхность пластины, и ось трубки. Теория этого метода изложена в [1].

Однако, в [1-2] неправильно принято, что сила поверхностного натяжения, приложенная к поверхности трубки в месте выхода ее из жидкости, пропорциональна смещению $x$ пластинки из положения равновесия. Если диаметр трубки постоянен вдоль ее длины, то эта сила постоянна и поэтому не входит в уравнение движения пластинки.

Кроме того, в [1] неправильно определена присоединенная масса жидкости на единицу площади – она в [1] завышена в два раза. Как видно из формулы, приведенной после формулы (24,6) на странице 123 в [4], сила трения на единицу площади колеблющейся по гармоническому закону плоскости (пластины) равна

$$-\sqrt{\eta\rho\omega/2} \cdot u(t) + \sqrt{\eta\rho\omega/2} \cdot du(t)/dt.$$

Влияние конечности размеров пластины на силу трения жидкости на пластинку учитывается через изменение площади пластинки на величину, равную произведению $H \cdot \delta$ высоты пластинки $H$ (размера в вертикальном направлении) на величину $2 \cdot \delta/2$, где $\delta = \sqrt{2\eta/\rho\omega}$, так как колебания происходят в вертикальной плоскости и пластинка имеет два вертикальных края и колебания происходят параллельно вертикальным краям. Это верно при $L \gg \delta$ и $H \gg \delta$. В [1] изменение площади неверно определено как произведение $L\delta/2$ ширины плоскости $L$ на величину $\delta/2$. Кроме того, в [1] часть силы трения, пропорциональная скорости $u(t)$, неправильно умножена на площадь двух сторон пластины $2S$, а часть силы трения, пропорциональная ускорению $du(t)/dt$ умножена на $2 \cdot S_{ef} = 2 \cdot (S + H \cdot \delta)$, в то время как обе эти части силы должны быть умножены на $2 \cdot S_{ef} = 2 \cdot (S + H \cdot \delta)$. Эти ошибки встречаются в [1] более десяти раз и во всех формулах, куда входит присоединенная масса. Эти же ошибки встречаются в [7-34] по истечении 36 лет со дня появления [1]. Присоединенная масса в [1] ни разу не приведена правильно, поэтому это не опечатка, а ошибка физическая, связанная с неграмотностью авторов [1,7]. Это тот случай, когда квалификация (подразумевающая наряду с другими наличие знаний - грамотности) ученого не соответствует его амбициям, что часто приводит к тому, что ученый становится маньяком.

В [1] не учтена сила Архимеда, пропорциональная смещению пластинки из состояния равновесия.

## Механические свойства колебательной системы [1].

Трубка из нержавеющей стали длиной 200мм диаметром 5мм висит на горизонтальных растяжках, сделанных из нержавеющей стали диаметром 0,3мм. По четыре растяжки верхнюю часть и середину трубки соединяют с цилиндрическим корпусом, при этом очень трудно добиться того, чтобы ось трубки оставалась строго вертикальной и чтобы растяжки находились в двух вертикальных плоскостях, линия пересечения которых совпадает с осью трубки. Поэтому очень трудно добиться, чтобы колебания системы, амплитуда которых составляет несколько микрометров, оставались в вертикальной плоскости. Это означает, что колебания не могут быть одномерными, отвечающими колебаниям вдоль оси трубки.

Диаметр цилиндрического корпуса не менее 30мм [1]. Поэтому длина растяжки не менее 12,5 мм. Для оценок примем, что длина растяжки равна 20мм, т.е. $l=20мм$. Легко показать, что амплитуде $b$ вертикальных колебаний системы отвечает удлинение растяжки, равное $\Delta l = b^2/2l$. Для $b=2мкм$ имеем $\Delta l = 10^{-7} мм = 0,1 мкм$. Для того, чтобы точность определения амплитуды была 1%, необходимо, чтобы длина каждой из 8 растяжек была определена с точностью до 0,02 мкм=20нм. Очевидно, что достичь такой точности для длины растяжки невозможно, так как диаметр растяжки 0,3мм и размер крепления растяжки к трубке и корпусу не меньше. Это означает, что точность определения длины растяжки порядка 0,3мм – диаметра растяжки. Тем более практически невозможно добиться того, чтобы длины всех 8 растяжек были определены с указанной выше точностью. Поэтому нельзя добиться того, чтобы эти восемь растяжек вели себя как пружина с жесткостью $k$ для колебаний с амплитудой в несколько микрометров: неточность в длине одной растяжки порядка 0,3мм приводит к изменению коэффициента жесткости $k$ упругого элемента из 8 растяжек (пружины) на $k/8$. Авторы [1-3] не доказали, что растяжки ведут себя как одна пружина с неизменным коэффициентом жесткости, и они не исследовали этот вопрос. Вопрос здесь не в том, является упругий элемент линейным, а в том, что является ли коэффициент упругости постоянным.

Кроме того, легко показать, что не исключены крутильные колебания системы вокруг оси трубки, так как невозможно трубку зафиксировать так, чтобы она могла двигаться только вдоль вертикальной оси в плоскости, проходящей через всю плоскую часть пластинки и ось трубки. Крутильные колебания и вертикальные колебания трубки сильно связаны и невозможно отделить их друг от друга. Легко показать, что коэффициенты упругости для крутильных колебаний и вертикальных колебаний имеют одинаковый порядок величины. Отсутствие крутильных колебаний и маятникоподобных колебаний в системе авторы [1-3] не доказали, даже не пытались исследовать этот вопрос.

С помощью 8 растяжек невозможно добиться того, чтобы колебания центра масс колебательной системы происходили вдоль неподвижной вертикальной оси, так чтобы центр масс оставался на этой оси. Это означает, что невозможно добиться того, чтобы колебания были одномерными.

### Электромагнитные и электромеханические свойства системы для определения вязкости жидкости, использованной в [1]

Из рисунка 11 в [1] видно, что два одинаковых магнита длиной 15мм, впрессованные в трубку, имеют длину, примерно в два раза меньшую длины катушки для возбуждения колебаний (активная катушка). Длина катушки для регистрации колебаний (пассивной катушки) равна длине активной катушки, находящейся ниже пассивной катушки. Магниты находятся внутри катушек. Расстояние между центрами магнитов 70мм. Следовательно, расстояние между нижним краем пассивной катушки и верхним краем активной катушки равно 70мм-15мм-15мм=40мм. Магнитное поле, создаваемое активной катушкой, через металлическую трубку и воздушное пространство внутри трубки проникает в пассивную катушку. Две катушки и трубка, проходящая через общую ось катушек, образует трансформатор. Пассивная трубка регистрирует сумму этого поля и магнитного поля колеблющегося магнита внутри нее. Эти два поля имеют одинаковую частоту и разные фазы, если считать колебания одномерными и удовлетворяющими принятым в [1] предположениям. Ток (сигнал), возбужденный в пассивной катушке, будет характеризовать только колебания системы, если проникшее поле будет много меньше магнитного поля магнита внутри пассивной катушки. В общем случае сигнал является результатом действия непостоянного в пространстве проникшего магнитного поля и магнитного поля магнита внутри пассивной катушки, являющегося непостоянным в пространстве, так как длина магнита меньше длины катушки. Авторы [1] не приняли никаких мер для изучения этого вопроса.

Периодические колебания электрического тока, текущего в активной катушке, могут быть и не гармоническими. Не проведена проверка гармоничности этих колебаний. Поэтому сила, действующая на магнит внутри активной катушки, может быть не гармонической.

Сила, действующая на магнит внутри активной катушки, может быть не гармонической в силу того, что магнитное поле, создаваемое катушкой, непостоянно в пространстве, даже если ток в катушке изменяется во времени по гармоническому закону. Поэтому нужно изучить вынуждающую силу, действующую на трубку, и добиться того, чтобы она была гармонической. При этом проверку гармоничности силы нужно провести не по сигналу, полученному от пассивной катушки, по причине, изложенной выше. Таким же способом нужно провести независимую проверку гармоничности колебаний пластинки, жестко связанной с трубкой.

Под действием магнитного поля, создаваемого активной катушкой, в металлической трубке создаются электрический ток, текущий вокруг оси трубки. В силу закона сохранения момента импульса, трубка должна вращаться в направлении, противоположном направлению электронов в токе. Поскольку поле периодическое, то и вращение тоже периодическое – трубка совершает крутильные колебания.

Трубка частично погружена в жидкость, поэтому магнитное поле, проникшее в жидкость, может привести к возникновению токов в проводящей жидкости, например в расплавах металлов и их сплавов. Это тоже может способствовать возникновению крутильных колебаний.

Экспериментальные установки [1-3] для измерения вязкости и проведения фазового анализа одно- и многокомпонентных систем, методы определения вязкости, методы определения плавления и затвердевания, разработаны на основе теории [1]. Эта теория, как показано выше, неверна. Кроме того, никем не проведена опытная проверка того, что экспериментальная установка описывается уравнением движения, изложенным в [1], вынуждающая сила изменяется по гармоническому закону и колебания пластинки являются гармоническими. Поэтому все результаты, полученные на основе вибрационного метода, и сам метод требуют тщательной проверки.

Очевидно, что в случае, когда жидкость не смачивает пластинку (угол смачивания равен 180 градусам), скорость жидкости возле пластинки не равна скорости самой пластинки, и поэтому нельзя использовать силу трения и присоединенную массу, полученную в [4]. Это же верно и в том случае, когда углы смачивания пластинки слабосвязанными между собой компонентами жидкости существенно отличаются. Поэтому мы в дальнейшем считаем, что угол смачивания меньше 180 градусов.

Кроме того, сомнительно применение характеристик вискозиметра, основанного на вибрационном методе, определенных для одного эталонного вещества, для других веществ с другим углом смачивания пластинки.

Осаждение на пластинке и трубке примесей, имеющихся в жидкости, может привести к изменению колебательных характеристик системы, что приводит к сбою калибровки системы и дает непредсказуемые по величине ошибки.

Как отмечено в [6, стр.102], пока краевой угол, образуемый пластинкой (трубкой), и кристаллом новой фазы (зародышем), образуемым на пластинке, меньше 180 градусов, работа образования зародыша на пластинке всегда меньше, чем работа образования зародыша внутри жидкости (расплава). Поэтому эти зародыши могут образоваться на пластинке даже в отсутствии примесей. Это также изменяет колебательные характеристики системы при температурах, гораздо выше температуры кристаллизации (затвердевания), что приводит к сбою калибровки системы и дает непредсказуемые по величине ошибки. Поскольку пересыщение возможно, лишь если работа образования зародыша больше нуля, то выйти за пределы условий равновесного существования фазы вблизи пластинки невозможно [6], то есть переход жидкости в метастабильную область невозможен.

Почти каждое происходящее в природе или лаборатории образование новых - жидких или газообразных фаз при малых или умеренных отклонениях от равновесия происходит на границах раздела, и всегда следует принимать специальные меры предосторожности, чтобы предотвратить эти явления и обеспечить возможность наблюдения процесса гомогенного образования зародышей [6,стр.103]. В рассматриваемой колебательной системе пластинка и трубка должны быть, и они образуют границу раздела между твердой и жидкой фазами, поэтому в этой системе мы не можем перейти к метастабильной жидкости.

Поэтому с помощью этой колебательной системы невозможно изучение процесса затвердевания и плавления одно- и многокомпонентных веществ. Это означает, что вибрационный метод фазового анализа, предложенный в [3,7-34], не является таковым. В связи с этим становятся очень сомнительными результаты и выводы многочисленных из работ [3,7-34], где якобы изучены процессы затвердевания и плавления с переходом к метастабильной жидкости. На наш взгляд, эти «Экспериментальные» результаты вымышленные или подогнаны под уже известные. Доказательства обнаружения новых фаз некоторых расплавов (смесей) веществ, полученные в некоторых из [3,7-34], также весьма сомнительны. Поэтому метод вибрационного фазового анализа неосуществим физически и у этого метода нет никаких перспектив, нет у него также ни настоящего и ни прошлого.

Если температура исследуемой жидкости высокая, то блок катушек и часть трубки, находящаяся в катушках, и упругий элемент должны иметь комнатную температуру, для того, чтобы можно было считать, что свойства упругого элемента и калибровку считать неизменными. Поэтому для проведения исследований по изучению вязкости и фазового анализа верхняя часть трубки (до точки входа в измерительный датчик) охлаждается до комнатной температуры водой, которая обтекает трубку с некоторой скоростью, зависящей от расхода воды. Причем тем больше температура в исследуемой жидкости, тем больше расход воды. Поэтому коэффициент сопротивления колебательной системы зависит от расхода воды, ее вязкости, от температуры в исследуемой жидкости. Поэтому не может быть и речи о постоянстве характеристик колебательной системы.

Кроме того, в системе возникают градиент температуры, по порядку величины равный $\frac{T-300}{15} \frac{K}{см}$, изменяющийся от 30 К/см до 67 К/см при изменении температуры исследуемой жидкости $T$ от 750$K$ до 1300$K$. Поэтому термодинамического равновесия в жидкости и в системе нет, и не может идти речь об измерении вязкости в зависимости от температуры, проведении фазового анализа равновесия жидкость - твердое тело и переходе жидкости в метастабильное состояние.

В фазовом равновесии в системе участвует, кроме веществ, из которых состоит исследуемая жидкость (далее компоненты), вещество (муллит или платина), из которого сделан сосуд, в котором находится исследуемая жидкость, вещество (нержавеющая сталь, вольфрам, платина), из которого сделана труба, и вещество (платина), из которого сделана пластина-зонд.

В некоторых случаях во избежание ликвации проводился барботаж жидкости (расплава) воздухом (или иным газом) в фазовом равновесии участвуют еще и вещества, из которых состоит газ, вводимый жидкость вблизи пластины, а также вещество (платина), из которого сделана часть, вводимая в жидкость, трубки для подачи газа. Поэтому полученные этим методом характеристики фазового равновесия относятся ко всей системы, состоящей из множества веществ, а не только из веществ, составляющих жидкость, если даже удастся добиться равновесия в системе, и эти характеристики не могут относиться к свойствам жидкости. В работах [1-3, 7-34] полученные характеристики выдаются как свойства жидкости. Это свидетельствует о том, что сообщаемые в этих работах результаты не были получены этим же методом, а являются результатом фальсификации, так как они совпадают с реальными данными для жидкости, полученными другими авторами другими бесконтактными методами. Поэтому, большинство результатов, полученных в [1-3, 7-34] фальшивые, следовательно не имеют никакой научной новизны и не являются научными результатами.

Кроме того, барботаж газа может привести к изменению вязкости, определяющей силу трения жидкости, действующей на пластинку.

В работе приведен далеко не полный перечень причин, из-за которых вибрационные методы измерения вязкости и фазового анализа вовсе не являются методами, или дают неверные результаты, или принципиально не могут дать то, что декларируется в работах [1-3, 7-34], или приводят к неконтролируемым ошибкам, или сильно увеличивают погрешности. Когда в «методе» есть неконтролируемые явления и необходимо создать трудноосуществимые условия, нужно отказаться от применения этого «метода» и лучше пользоваться другими методами. Поэтому лучше не применять вибрационные методы измерения вязкости и фазового анализа, предложенные и использованные в [1-3, 7-34], и не пользоваться результатами, полученными в [1-3, 7-34].

## Список литературы